\documentstyle[12pt]{article}

\makeatletter \@addtoreset{equation}{section}

\newcommand{\be}{\begin{equation}}
\newcommand{\ee}{\end{equation}}
\newcommand{\bear}{\begin{eqnarray}}
\newcommand{\eear}{\end{eqnarray}}
\newcommand{\ba}{\begin{array}}
\newcommand{\ea}{\end{array}}

\makeatletter \@addtoreset{equation}{section}

\makeatother

\begin{document}

\begin{titlepage}
\vfill
\begin{flushright}
{\normalsize KIAS-P04005}\\
{\normalsize hep-th/0401078}\\
\end{flushright}

\begin{center}

\vfill

{\Large\bf A Holographic View on\\ Matrix Model of Black Hole}

\vspace{0.3in}

Takao Suyama \footnote{e-mail address : suyama@kias.re.kr} and
Piljin Yi \footnote{e-mail address :
piljin@kias.re.kr}

\vspace{0.15in}

{\it School of Physics, Korea Institute for Advanced Study,}

{\it 207-43, Cheongnyangni 2-dong, Dongdaemun-gu, Seoul 130-722, Korea}

\vfill

\end{center}

\begin{abstract}
\noindent We investigate a deformed matrix model proposed by
Kazakov et.al. in relation to Witten's two-dimensional black hole.
The existing conjectures assert the equivalence of the two by
mapping each to a deformed $c=1$ theory called the sine-Liouville
theory. We point out that the matrix theory in question may be
naturally interpreted as a gauged quantum mechanics deformed by
insertion of an exponentiated Wilson loop operator, which gives us
more direct and holographic map between the two sides. The matrix
model in the usual scaling limit must correspond to the bosonic
$SL(2,R)/U(1)$ theory in genus expansion but exact in $\alpha'$.
We successfully test this by computing the Wilson loop expectation
value and comparing it against the bulk computation. For the
latter, we employ the $\alpha'$-exact geometry proposed by
Dijkgraaf, Verlinde, and Verlinde, which was further advocated by
Tseytlin. We close with comments on open problems.

\end{abstract}

\vfill

\end{titlepage}

\newpage

\section{Introduction}

Recently, the matrix model of two dimensional noncritical string
theory came under spotlight in a new context. When the matrix
model was incepted some fifteen years ago,\footnote{See
Ref.~\cite{Ginsparg} for a general review of the original matrix
model proposal.} it started as a theory of triangulated worldsheet
whose structure is encoded in the Feynman diagram of the quantum
mechanics in question. Under the new interpretation, however, the
matrix is more than such an auxiliary field. It is reinterpreted
as open string tachyons living on many unstable D-branes, whose
condensation leads to the two dimensional string theory
\cite{reloaded}. In this new interpretation, restriction to the
singlet sector becomes quite natural since the open string side
typically has gauge symmetry.

On the other hand, it is  well-known that the two-dimensional
string theories admit black hole backgrounds
\cite{2dimBH1}\cite{2dimBH2}\cite{2dimBH3}. This background was
first considered by Witten, as a coset model of $SL(2,R)/U(1)$
type. Relationship among this theory, usual $c=1$ noncritical
string theories, and matrix models has been the subject of a good
deal of interests since early 1990's. For the case with ${\cal
N}=$(2,2) supersymmetries on the worldsheet, there is a strong
evidence that the black hole theory is in fact the same as the
super-Liouville theory \cite{Hori}. For cases with lesser
supersymmetries, however, the situation is slightly different. For
instance, in the bosonic case, the black hole theory is
conjectured by Fateev, Zamolodchikov, and Zamolodchikov (FZZ)
\cite{FZZ} to be reached by perturbing usual Liouville theory with
condensation of vortices, leading to so-called sine-Liouville
theory.

Recently, a matrix model is proposed by Kazakov, Kostov, and
Kutasov in \cite{KKK} to be dual to this sine-Liouville theory,
which would then also imply that the matrix model will describe
the black hole theory upon using the FZZ conjecture. This matrix
model can be regarded as a deformation of the $c=1$ matrix model.
Their calculation of the matrix model  supports the equivalence
between the matrix model and the sine-Liouville theory. Although
the proposal is convincing, a more direct relation between the
matrix model and the black hole would be desirable.

In this paper we employ a holographic view point on this matrix
model, and show a rather direct map to the black hole. Our
demonstration is based on the observation that the matrix model in
\cite{KKK} is actually a gauge theory. Then, in analogy with the
well-known AdS/CFT correspondence \cite{ADS-CFT}, it is natural
to assume that the expectation value of Wilson loop operator is
calculated from a partition function of a string whose boundary is
fixed on a circle on which the Wilson loop is defined. Thus, from
this observable one can extract some geometric data for the target
spacetime of the holographically-dual string theory. We will show
that the result obtained from the gauge theory side nicely fits
with the one obtained from the $\alpha'$-exact string background
\cite{exactBGD1}\cite{exactBGD2}.

This paper is organized as follows. Section \ref{rev} contains a
brief review on the matrix model. In section \ref{holo}, we
explain that the matrix model is actually a gauge theory, then
discuss the holographic description of the matrix model. In
section \ref{Wilson} we compute the Wilson line expectation value
in the matrix side and compare it to the expected results from the
proposed closed string dual, namely the two dimensional black
hole. Section \ref{dis} is devoted to discussion.

\section{Review of a matrix model of two-dimensional black hole} \label{rev}

In this section, we briefly review the proposal of \cite{KKK} on a
matrix model which describes a string theory defined on a
two-dimensional black hole. See also \cite{review} for a
comprehensive review.

Consider the $c=1$ matrix quantum mechanics compactified on a
circle with radius $R$ with a twisted boundary condition, whose
partition function is
\begin{equation}
Z_N(\Omega) = \int_{M(2\pi R)=\Omega^\dag M(0)\Omega}{\cal D}M(x)\ \exp\left(-\mbox{Tr}\int_0^{2\pi R}
              dx\left[\frac12(\partial_xM)^2+V(M)\right]\right),
\end{equation}
where $M(x)$ is an $N\times N$ matrix-valued field on the circle and $\Omega$
is a unitary matrix. The form of $V(M)$ is
\begin{equation}
V(M) = \frac12M^2-\frac g{3\sqrt{N}}M^3,
\end{equation}
but the precise form is not important in the double scaling limit.
The proposed matrix model is defined by the following partition
function
\begin{equation}
Z_N(\lambda) = \int D\Omega\ e^{\lambda\mbox{\small Tr}(\Omega+\Omega^\dag)}Z_N(\Omega).
   \label{KKKmodel}
\end{equation}
Note that this would be identical to the partition function of the
Gross-Witten model \cite{GrossWitten} if we drop $Z_N(\Omega)$ on
the right hand side. In this sense, $Z_N(\lambda)$ represents a
coupled theory of Gross-Witten model and $c=1$ matrix model.

The claim of Ref.~\cite{KKK} is that the matrix model
(\ref{KKKmodel}) describes in the double scaling limit a string
theory defined on a two-dimensional black hole with temperature
$T=1/2\pi R$. This claim is based on the FZZ conjecture \cite{FZZ}
which shows some pieces of evidence for an equivalence between
$SL(2,R)/U(1)$ coset CFT with level $k$, which is known to be a
CFT description of the two-dimensional black hole
\cite{2dimBH1}\cite{2dimBH2}\cite{2dimBH3} when $k=9/4$, and the
sine-Liouville theory
\begin{equation}
S = \frac1{4\pi}\int d^2\sigma\left[ (\partial x)^2+(\partial\phi)^2-\frac1{\sqrt{k-2}}R^{(2)}\phi
   +\lambda e^{-\sqrt{k-2}\phi}\cos \sqrt{k}(x_L-x_R) \right].
\end{equation}
See Ref.~\cite{KKK} for complete detail. There it is shown that
the matrix model (\ref{KKKmodel}) is equivalent to a modified
version of the sine-Liouville theory
\begin{equation}
S = \frac1{4\pi}\int d^2\sigma\left[ (\partial x)^2+(\partial\phi)^2-2R^{(2)}\phi
   +\mu e^{-2\phi}+\lambda e^{(R-2)\phi}\cos R(x_L-x_R) \right].
      \label{modifiedSL}
\end{equation}
In a suitable limit, in which the term $\mu e^{-2\phi}$ is
negligible, and with $R=3/2$, this modified theory reduces to the
sine-Liouville theory with $k=9/4$. Evidence for this equivalence
is provided  by calculating the free energy of the matrix model,
and the results show exact $\lambda$-dependence of the free energy
expected from the worldsheet theory (\ref{modifiedSL}). Thus, by
combining these two claims, it is proposed that the matrix model
(\ref{KKKmodel}) with $R=3/2$ in a suitable limit would be
equivalent to a string theory whose tree level dynamics is
governed by $SL(2,R)/U(1)$ coset CFT with $k=9/4$. Note that
varying $R$ in the matrix model is not equivalent to varying $k$
in the coset CFT. Varying $R$ corresponds to varying the
temperature, and this enables  discussion of a thermodynamical
issue on the black hole.

In the course of computation (\ref{KKKmodel}), the matrix $M$ is
integrated out first, and thus we cannot reduce the system to that
of eigenvalues of $M$. This forces us to employ a method different
from the ordinary matrix model in taking the double scaling limit.
Instead of introducing Fermi level in canonical ensemble,
Ref.~\cite{KKK} introduced  the grand canonical partition function
\begin{equation}
Z(\lambda,\mu) = \sum_{N=0}^\infty e^{2\pi R\mu N}Z_N(\Lambda) ,
   \label{GCPF}
\end{equation}
where the chemical potential $\mu$ takes the role of the effective
Fermi level. It has been shown in \cite{grandcanonical} that the
$1/\mu$ expansion of $F(\lambda,\mu)\equiv\log Z(\lambda,\mu)$
reproduces the genus expansion of $\log Z_N(\lambda)$ in the
double scaling limit. The introduction of the grand canonical
partition function appears to be very powerful method since
$Z(\lambda,\mu)$ is the $\tau$-function of an integrable system,
and as a result, $Z(\lambda,\mu)$ can be obtained by solving a
partial differential equation.

It should be noted that the matrix model should be defined with a
{\it finite} cut-off $\Lambda$ which determines the range of
integration of $M$. In \cite{KKK}, it is proposed that the black
hole is made of non-singlet states of $SU(N)$ in the matrix model,
while the non-singlet states become infinitely heavy compared with
the singlet states as $\Lambda\to\infty$ \cite{cutoff}. Thus one
should keep $\Lambda$ finite as long as one would like to obtain a
black hole.

\section{Black hole matrix model as a gauge theory and holography} \label{holo}

A main reason for recent resurgence of matrix model is that the
latter found a new and more appealing interpretation as an open
string theory of unstable D-branes. This endowed the matrix model
of $c=1$ strings with a strong flavor of open string / closed
string duality, along  the line of AdS/CFT correspondence. In
describing the matrix model as an open string theory, the gauging
is conceptually important since, the gauge field is a universal
aspect of D-brane, much as gravity is universal to critical closed
string theory. In  light of this, one may ask whether the correct
matrix model of bosonic black hole in two dimensions should be
also a gauged quantum mechanics. Here we wish to point out that
the matrix model proposed by Kazakov et.al. is itself a gauge
theory as it is.

For this, let us introduce $U(N)$ gauge field $A$, again $N\times N$
matrix, and let $M_0$ be
in the adjoint representation under this  $U(N)$. We elevate the
original matrix model, prior to the deformation, to a gauged
version by replacing the time derivative of $M_0$ by a covariant one
\begin{equation}
\partial_x M_0(x) \quad\rightarrow\quad \partial_x\tilde M(x) +i[A(x),\tilde
M(x)],
\end{equation}
where we used a new notation $\tilde M$ to emphasize that it is
now coupled to the gauge field, and at the same time integrating
over the gauge field in the path integral;
\begin{equation}
\int {\cal D}M_0(x) \quad \rightarrow \quad \int {\cal D}A(x){\cal
D}\tilde M(x) .
\end{equation}
Usually this will simply impose the Gauss constraint and do
nothing else.

With periodic $x$, however, a new physical variable emerges from
the gauge sector in the form of the holonomy, which is nothing but
the untraced Wilson loop;
\begin{equation}
{\cal W}=P e^{i\oint A \,dx} .
\end{equation}
We may bring $A(x)$ to a constant matrix $\bar A$, using small
gauge transformations, but only in a manner that leaves the
holonomy invariant. Holonomy would change upon a non-single-valued
gauge transform but the latter is not a valid gauge
transformation. Thus, dividing by proper gauge volume, we find
reduction of the path integral as
\begin{equation}
\int {\cal D}A(x){\cal D}\tilde M(x) \quad \rightarrow \quad \int
D\bar A\,\int {\cal D}\tilde M(x) .
\end{equation}
It is important to note here that the scalar $M_0(x)$ must
satisfy periodic boundary condition, for we have not modified
anything other than introducing the gauge field.

Now it is also clear that we may use a large (i.e.
non-single-valued) gauge transformation,
\begin{equation}
U=e^{ix\bar A} ,
\end{equation}
to induce a "change of variable," upon which we find
\begin{equation}
\tilde M(x) \quad \Rightarrow \quad U(x)^\dagger \tilde M(x)U(x) ,
\end{equation}
while
\begin{equation}
\bar A \quad\Rightarrow \quad 0 .
\end{equation}
Since $\tilde M$ is a dummy variable of the path integral, the
only relevant point of this is that the transformed matrix $M'
\equiv U^\dagger\tilde M U$ obeys a modified periodicity condition
that
\begin{equation}
M'(2\pi R )= U(2\pi R)^\dagger M'(0) U(2\pi R) .
\end{equation}
By now, it should be pretty clear that we must identify variables
$M(x),\Omega$ of the  matrix model as
\begin{equation}
M=M',\qquad \Omega=U(2\pi R) ,
\end{equation}
and
\begin{equation}
\int D\bar A\,\int {\cal D}\tilde M(x)\quad = \quad \int D\Omega
\int_{M(2\pi R)=\Omega^\dag M(0)\Omega} {\cal D}M(x) .
\end{equation}
The final ingredient in making of the deformed matrix model is the
insertion of
\begin{equation}
e^{\lambda\mbox{\small Tr}(\Omega+\Omega^\dag)} ,
\end{equation}
which is nothing but the exponential of the traced Wilson
loop operator,
\begin{equation}
e^{\lambda \hbox{Tr} ({\cal W}+{\cal W}^\dagger)} .
\end{equation}
Thus we conclude that the deformed matrix model is in itself a
gauged quantum mechanics. One effect of this perturbation is to
modify the Gauss constraint, which otherwise would have truncated
non-singlet states altogether. With finite $\lambda$, The modified
Gauss constraint now allow contributions from non-singlet sector
with appropriate weight determined by the Wilson loop operator.

With this modest observation, we are ready to consider a more
direct comparison between the deformed matrix model and bosonic
black holes in two dimensions, by making use of familiar
holographic relations between gauge theories and closed string
theories. Let $Y$ denote the target spacetime of the dual string
theory. $Y$ is a two-dimensional Euclidean manifold since the
matrix model is supposed to describe a black hole in a thermal
background. Since the matrix model is defined on $S^1$, $Y$ should
be asymptotically ${\bf R}\times S^1$. $Y$ should also have a
$U(1)$ isometry reflecting the time-translation symmetry of the
matrix model. Thus there are only two possibilities for the
topology of $Y$; an infinitely long cylinder or an semi-infinite
cigar type geometry.

With the fact that the matrix model is equivalent to a gauge
theory perturbed by Wilson loop, we now have an order parameter to
determine the topology of $Y$, and to some extent its geometry as
well. Consider the expectation value of the traced Wilson loop
operator defined for the time circle,
\begin{equation}
W\equiv \frac1N\,\mbox{Tr}\,{\cal W} .
\end{equation}
In analogy with the well-known AdS/CFT
correspondence \cite{ADS-CFT}, $\langle W\rangle$ must be
identifiable with the partition function of the
dual string whose boundary is fixed on $S^1$, i.e.
\begin{equation}
\langle W\rangle =\left\langle \frac1N\,\mbox{Tr}Pe^{i\oint Adx}\right\rangle
= \int {\cal D}X\ e^{-S_{NG}},
   \label{def_of_W}
\end{equation}
where $S_{NG}$ is the Nambu-Goto action and the path integral
is over the worldsheets with the boundary wrapping the
asymptotic circle.

Since the dual string lives in a two-dimensional Eulidean
spacetime, call it $Y$, it has no fluctuating degrees of freedom.
This means that the path integral in the string side is evaluated
by calculating the area of the string wrapping on $Y$, or the area
of $Y$ itself. {}From the observation that the unperturbed matrix
model is dual to the $c=1$ Liouville theory, it is reasonable to
suppose that the shape of $Y$ is asymptotically a cylinder to one
side. As argued in \cite{KKK}, the matrix model is supposed to be
defined with a finite cut-off, if we wish to describe the
two-dimensional black hole. According to this, then, the boundary
theory should be placed not at the asymptotic boundary of $Y$ but
at some finite position. In other words, when we calculate the
area of the worldsheet, $Y$ should be regarded as either a
semi-infinite cylinder ${\bf R}_{\ge0}\times S^1$ or a cigar with
a finite length, depending on whether it corresponds to some
Liouville like theory or a black hole theory.

Then the strategy to determine the topology of the spacetime is
now clear. The order parameter $\langle W\rangle$ will vanish if
and only if the geometry is semi-infinite cylinder,
\begin{equation}
\langle W\rangle = 0 \hspace{3mm} \Leftrightarrow \hspace{3mm} Y\
\cong\ {\bf R}_{\ge0}\times S^1 ,
\end{equation}
while for cigar geometry, it should measure the area inside UV
cut-off. We will show below that $\langle W\rangle$ has a
non-vanishing value, indicating that the holographic dual string
theory indeed lives on a cigar like geometry. With this, it
becomes quite likely that the matrix model in question is the
holographic dual of a string theory on a two-dimensional black
hole.

In fact, one can also probe for  more detail of the background
geometry of the dual string theory by examining the value of
$\langle W\rangle$. In the closed string side, this expectation
value is computed by the saddle-point evaluation of string action,
which contains a factor of the exponential of minus the worldsheet
area. As we will see presently, this area contains a rather crude
information about the black hole, yet detailed enough to
distinguish between the leading $\alpha'$ order geometry and the
$\alpha'$-exact geometry. In next section, we will study this
order parameter in much detail.

\section{Wilson loop and $\alpha'$-exact dual geometry} \label{Wilson}

In this section we evaluate $\langle W\rangle$ in the  matrix
model, and also in its closed string dual. For the comparison, we
must recall that the precise duality is proposed for $c=26$ case
of bosonic black hole, which corresponds to $k=9/4$. In the matrix
side, this is supposed to match with $R=3/2$ \cite{KKK}. Below we
will see precise match of Wilson loop expectation value in terms
of its leading string coupling behavior and its UV cut-off
dependence.

\subsection{Wilson loop from the bulk}

In general, quantities derived from the matrix model may be
provided as a genus expansion. The contribution of each genus
would have all $\alpha'$ corrections since there is no expansion
parameter corresponding to $\alpha'$. This means that from the
matrix model side, as we will show later, we obtain a result which
should be correct to all orders in $\alpha'$. On the other hand,
the familiar black hole geometry
\begin{eqnarray}
ds^2 &=& \alpha'k\left[ dr^2+\tanh^2r\,d\theta^2\right]  ,\nonumber \\
\Phi &=& \Phi_0-\frac12\log(\cosh 2r+1) ,
\end{eqnarray}
is derived by considering the bare Lagrangian of gauged
WZW model of $SL(2,R)/U(1)$ and is not exact in $\alpha'$.
Even though we are considering only the area of this
black hole, which is a crude measure of the geometry,
higher order correction could be nontrivial for the simple
reason that the asymptotic circle size is of order
$\sqrt{k\alpha'}$. For finite $k=9/4$ that we are considering,
a string winding around this circle has mass of order string
scale, so the extended nature of the string become important.

Fortunately, an $\alpha'$-exact modification of this geometry has
been proposed by Dijkgraaf, Verlinde, and  Verlinde via study of
conformal field theory of the gauged WZW model \cite{exactBGD1}.
This was further advocated by Tseytlin \cite{exactBGD2} using the
following line of reasoning: The bare action of $G/H$ gauged WZW
theory may be written as
\begin{equation}
I_k(g,A_z,A_{\bar z})=k I_{WZW}(h^{-1}g\bar h) ,
-k I_{WZW}(h^{-1}\bar h)
\end{equation}
in terms of usual WZW action $I_{WZW}$.
Here, $A_z=h\partial h^{-1}$ and $A_{\bar z}=\bar h
\bar \partial \bar h^{-1}$. It was then argued that the
only correction is to shift the level $k$ in each term
giving us,
\begin{equation}
I_k^{exact}(g,A_z,A_{\bar z})=\left(k+c_G/2\right)I_{WZW}(h^{-1}g\bar h)
-\left(k+c_H/2\right) I_{WZW}(h^{-1}\bar h) ,
\end{equation}
with the dual Coxeter number $c_G$ and $c_H$ of the respective
gauge groups. One may extract the background by integrating out
the gauge field as usual, and the resulting exact geometry
is\footnote{One unusual aspect of this $\alpha'$-corrected black
hole geometry is that it has no curvature singularity when
continued to the Minkowskian signature \cite{smooth}. The
Minkowskian geometry does have event horizon but no curvature
singularity inside. }
\begin{eqnarray}
ds^2 &=& \alpha'(k-2)
\left[ dr^2+\frac{\tanh^2r}{1-(2/k)\tanh^2r}d\theta^2 \right],
\nonumber \\
\Phi &=& \Phi_0-\frac14\log(\cosh 2r+1)
-\frac14\log\left(\cosh 2r+1+\frac4{k-2}\right).
\end{eqnarray}
Note that the corrections to the dilaton and to the radius of the
$\theta$-direction is exponentially small in the asymptotic
region, while the correction to the radial coordinate is more
prominent due to $k\rightarrow (k-2)$ in the overall factor. This
means among other things, the area bound by a particular UV value
of $\Phi$ is different before and after the $\alpha'$ correction.

The value of the Nambu-Goto action is simply the area divided by
$2\pi\alpha'$. The area $A(r_0)$ for a region $r\le r_0$ is
\begin{eqnarray}
A(r_0)
&=& 2\pi\alpha'\sqrt{k(k-2)}\left[
    \log\left(\cosh r_0+\sqrt{\cosh^2r_0+\frac2{k-2}}\right)\right. \nonumber \\
& & \left. -\log\left(1+\sqrt{1+\frac2{k-2}}\right)\right] \nonumber \\
&=& 2\pi\alpha'\sqrt{k(k-2)}r_0+O(1).
\end{eqnarray}
For large $r_0$, dilaton $\Phi$ behaves as
\begin{equation}
\Phi(r_0) - \Phi(0) = -r_0+O(1).
\end{equation}
Thus we conclude that
\begin{equation}
e^{-A/(2\pi\alpha')} \sim \exp\left(-\sqrt{k(k-2)}(\Phi(0)-\Phi(r_0))\right) .
\end{equation}
Note that the leading $\alpha'$ geometry
 \cite{2dimBH1}\cite{2dimBH2}\cite{2dimBH3} would have produced
 $\sqrt{k^2}$ in place of $\sqrt{k(k-2)}$. With $k=9/4$, the two
 differs by a factor of 3, which  means that the equivalence
between the matrix model and string theory on the two-dimensional
black hole can be tested to all orders of $\alpha'$ by examining the coefficient.

To summarize, we evaluated the Wilson loop expectation value of
the open string side in terms of its proposed dual in the closed
string side and found
\begin{equation}
\langle W \rangle \sim
\exp\left(-\sqrt{k(k-2)}(\Phi(0)-\Phi(r_0))\right) ,
\end{equation}
where $\Phi(r_0)$ is the value of dilaton where the Wilson line,
or the boundary of the fundamental string is located while
$g_{st}\equiv \exp(\Phi(0))$ is the value of string coupling at
the horizon. Presently we will see that with $k=9/4$ for which the
dual pair is proposed, the coefficient $\sqrt{k(k-2)}=3/4$ matches
the result of the matrix model computation exactly.

\subsection{Wilson loop from the matrix model}

Now we perform the matrix model calculation.
As has been shown previously, the twist matrix $\Omega$
is a holonomy along the time circle. Thus the Wilson loop operator is
\begin{equation}
W =\frac1N\mbox{Tr}\ {\cal W}= \frac1N\mbox{Tr}\ \Omega.
\end{equation}
Employing the grand canonical method as in \cite{KKK},
we will compute,
\begin{equation}
\langle W\rangle
\equiv \frac1{Z(\lambda,\mu)}\sum_{N=0}^\infty e^{2\pi R\mu N}\int D\Omega {\cal D}M\
   e^{\lambda \mbox{\small Tr}(\Omega+\Omega^\dag)-S(M)}\ \frac1N\mbox{Tr}\ \Omega,
\end{equation}
where $Z(\lambda,\mu)$ is the grand canonical partition function
(\ref{GCPF}).

It is easy to show that $\langle W\rangle$ is a solution of the
following differential equation
\begin{equation}
\partial_\mu\langle W\rangle + \partial_\mu\log Z(\lambda,\mu)\cdot\langle W\rangle
= \pi R\partial_\lambda\log Z(\lambda,\mu).
    \label{W}
\end{equation}
In the limit $\lambda\to\infty$ while $\mu$ fixed, which is the
proposed limit for the matrix model to be equivalent to the
two-dimensional black hole, $F(\lambda,\mu)=\log Z(\lambda,\mu)$
behaves as
\begin{equation}
F(\lambda,\mu) = -A\lambda^{4/(2-R)}-B\mu\lambda^{2/(2-R)}+O(\lambda^0),
   \label{FE}
\end{equation}
where $A,B$ depend only on $R$. In Ref.~\cite{KKK}, the sign of
$B$ here appears to be ambiguous in the computation. The free
energy is obtained not by explicit path integral, but rather by
solving a second order differential equation which is even with
respect to $\mu$. We will fix this sign so that the expectation
value of $N$ comes out to be positive, which is sensible in the
spirit of original matrix model proposal. This sign choice is on
par with that of Ref.~\cite{kazakov}.

In this limit, the equation (\ref{W}) becomes
\begin{equation}
\partial_\mu\langle W\rangle - B\lambda^{2/(2-R)}\langle W\rangle
= -\frac{4\pi RA}{2-R}\ \lambda^{(2+R)/(2-R)}.
\end{equation}
If we ignore the derivative term, to which we will come back below,
this equation has a solution
\begin{equation}
\langle W\rangle \sim \lambda^{R/(2-R)}.
\end{equation}
Since $\lambda^{-2/(2-R)}\sim g_{st}$ \cite{KKK}, we finally obtain
\begin{equation}
\langle W\rangle \sim \exp\left(-\frac R2\log g_{st}\right).
   \label{result0}
\end{equation}
The above calculation is a bit naive, since this expression can be
arbitrary large for large $\lambda$ while $|\langle W\rangle|\le
1$ since the operator inside is a sum over $N$ complex numbers of
unit modulus, divided by $N$. This would be due to the fact that
 only terms independent of the cut-off $\Lambda$ are
derived. Thus the problem would be solved by taking into account
the cut-off $\Lambda$ dependence of the matrix model explicitly.
To see this, let us consider the worldsheet theory
\begin{equation}
S = \int d^2\sigma \frac1{4\pi}\left[ (\partial x)^2+(\partial\phi)^2-2R^{(2)}\phi
   +\mu e^{-2\phi}+\lambda e^{(R-2)\phi}\cos R(x_L-x_R) \right],
\end{equation}
which is equivalent to the deformed matrix model in question. The
cut-off $\Lambda$ is related to the maximum value $\phi_{UV}$ of
the Liouville field $\phi$. Thus the $\Lambda$-dependence can be
included in $\mu,\lambda$ by the scaling
\begin{equation}
\mu \to e^{-2\phi_{UV}}\mu, \hspace{5mm} \lambda\to e^{(R-2)\phi_{UV}}\lambda.
\end{equation}
Note that $\mu\lambda^{-2/(2-R)}$ is invariant under this scaling.
Note also that the free energy is also rescaled, but it is not
relevant for calculations below. This scaling results in the
scaling of the Wilson loop
\begin{equation}
\langle W\rangle \sim e^{-R\phi_{UV}}\exp\left(-\frac R2\log g_{st}\right).
\end{equation}
Note that this multiplicative factor is relevant since the
differential equation (\ref{W}) is inhomogeneous and it should be
determined irrespective of a boundary condition for the solution.
In the asymptotic region, the dilaton $\Phi$ behaves as
$\Phi\sim-2\phi$. Thus we obtain
\begin{equation}
\langle W\rangle \sim \exp\left(-\frac R2(\Phi-\Phi_{UV})\right).
   \label{result}
\end{equation}
This result (\ref{result}) indicates that, according to the
holographic interpretation discussed previously, the target
spacetime $Y$ has a cigar like geometry since it is non-vanishing.
Moreover, since the area $A$ read off from $\langle W\rangle$ is
\begin{equation}
A /2\pi\alpha'\sim \frac R2(\Phi-\Phi_{UV}),
\end{equation}
one can see that there is a linear dilaton background in the
asymptotic region. With $R=3/2$, this matches the closed string
result above precisely since the corresponding value of $k$ is
such that the coefficient of the exponent is $3/4$, upon
identifying $\Phi=\Phi(0)$ and $\Phi_{UV}=\Phi(r_0)$.

Curiously, solutions of the differential equation (\ref{W}) can
have a term
\begin{equation}
\exp\left(-2\pi R\int^\mu d\mu'\langle N\rangle \right),
   \label{NP}
\end{equation}
which is the homogeneous solution of Eq.~(\ref{W}). We have used
the fact that $\partial_\mu F(\lambda,\mu)=2\pi R\langle
N\rangle$. Since $\langle N\rangle\sim g_{st}^{-1}$, this term
looks like a non-perturbative correction to the Wilson loop.
However, it is not clear to us whether this piece is physical or
an artifact of taking the grand canonical ensemble. The term
arises out of correlation between $1/N$ and Tr$\ \Omega$, and
would be absent if we computed in the canonical ensemble with a
fixed large $N$.

Another fine detail overlooked above is that this matrix model
side computation gives the negative sign for $\langle W\rangle$
with the current sign choice such that $\langle N\rangle>0$,
unlike the bulk computation. However, $\Omega$ acts the matrix $M$
in the adjoint representation so that its sign cannot be fixed
unambiguously. We believe the sign ambiguity originates here and
should be chosen as necessary.

\section{Discussion} \label{dis}

We have investigated the matrix model of Ref.~\cite{KKK}, which is
proposed to be dual to sine-Liouville theory and, via FZZ
conjecture, thus also dual to bosonic string theory in the two
dimensional black hole background. We started with the observation
that this matrix model is actually the gauged version of usual
matrix model for two dimensional noncritical string theory, with a
crucial deformation introduced by insertion of exponentiated
Wilson loop for the time circle.

Then we argued that expectation value of the Wilson loop $\langle
W\rangle$ is an order parameter that determines the topology of
the Euclidean spacetime of the dual string theory. This is based
on usual relationship in open/closed duality of AdS/CFT type
situation, where Wilson line of the boundary theory is represented
by macroscopic open string with boundary at asymptotic region.  In
fact, $\langle W\rangle$ possesses more information than the
topology, since it measures the area bound by the Wilson loop.

We showed that $\langle W\rangle$ does not vanish, confirming the
cigar-like geometry of the black hole background. Furthermore we
found that the Wilson loop computation reproduces exactly the right
area dependency expected from the bulk side computation using
Nambu-Goto action, confirming in part the conjecture that this
matrix model is dual to string theory in black hole background. In
comparing the two sides, a crucial ingredient is the
$\alpha'$-exact black hole background previously given in
Ref.~\cite{exactBGD1, exactBGD2}.

A salient point of our result is that we can extract some {\it
geometric} data on the bulk side from the matrix model, which is
in general very difficult task to realize. This is due to the
simplicity of two dimensional theories. Another unusual aspect of
the current computation is that it is sensitive to $\alpha'$
correction to all orders,  since the matrix theory computation
should correspond to the leading order result in $g_{st}$ but the
exact result in $\alpha'$. In a sense, this is also a consistency
check for the $\alpha'$-exact black hole background, proposed some
dozen years ago.

Let us recall that Ref.~\cite{KKK} argues the equivalence between
the two sides, mainly by computing the partition function of the
matrix model and comparing it to that of sine-Liouville theory.
Equivalence to the black hole theory is then via another
conjecture by FZZ linking the sine-Liouville theory and the black
hole theory. In the present paper, we tried to give more direct
connection between the matrix model and black hole. While we
borrowed heavily from computations in \cite{KKK}, we do not rely
on the FZZ conjecture at all. Rather we made use of matrix theory
partition function to map observables in the matrix model to those
in string theory in the black hole background. In this sense, our
computation can be taken as a strong supporting evidence for the
FZZ conjecture, when taken together with main results of
Ref.~\cite{KKK}.

Let us close with some open questions. It is sometimes argued that
the density of states of the two-dimensional black hole exhibits
the Hagedorn behavior. This is simply based on the special form of
Bekenstein-Hawking entropy which has the form the black hole mass
divided by the constant Hawking temperature of the black hole. The
latter is then referred to as Hagedorn temperature of the system.
On the other hand, it is well-known that the singlet sector of the
$c=1$ matrix model has only one (continuous) degree of freedom,
namely the tachyon. It has been proposed that non-singlet states
would contribute to the Hagedorn behavior.

In our view point, the matrix model in question is actually a
gauge theory. Thus its physical Hilbert space has to be
constrained by the Gauss constraint. While the deformation
introduced by the Wilson loop insertion modifies the Gauss
constraint nonlinearly, allowing contribution from non-singlet
sector as well as from singlet sectors, it is unclear whether such
operation can change the net number of physical states. This seems
to suggest that the real physical degrees of freedom of the matrix
model can be much smaller than those expected based on Hagedorn
behavior of the black hole geometry. It is important to clarify
this issue. Possibly one relevant fact to consider is that the
deformation in question involves inserting infinite number of
Wilson loops, or in other word, inserting infinite number of
quark-antiquark pairs. This is because we insert an exponentiated
Wilson loop.

Another interesting problem on the black hole side is the matter
of Hawking radiation. The matrix model is purported to be a
nonperturbative formulation of the two-dimensional string theory,
and thus must contain full answer to the old questions on quantum
evoluton of black holes. While we gained a lot of understanding
about black holes during the last decade, those were largely for
extremal and near extremal black hole with a BPS-like endpoint.
For instance, the question of how to describe a large uncharged
black hole is still an open question. The two-dimensional black
hole is probably the simplest system where black hole must radiate
away completely and be replaced by something else. We hope that
our holographic view point would be helpful for understanding this
fundamental problem.

Recently, more varieties of matrix models are proposed for
fermionic string theories in two dimensions. The matrix theory for
type 0B theory is identified in \cite{TT}\cite{allstar} as the
same old matrix model but with different Fermi sea configurations.
The Fermi sea corresponding to type 0B string is stable even
non-perturbatively, contrary to the bosonic case. Matrix theory
dual of type 0A is also proposed in the same work. Also gauged
version of Marinari-Parisi model \cite{Marinari} has been proposed
to be dual to ${\cal N}=2$ super  Liouville theory \cite{N=2}.

It goes without saying that much is needed to be clarified how the
bosonic case described here elevates to fermionic cases. In
particular, the case of type 0 must be addressed in the present
matrix theory context, since it simply corresponds to a different
choice of vacuum. Type II case is also very intriguing, given the
known mirror symmetry that asserts that ${\cal N}=2$ Liouville theory is
identical to ${\cal N}=2$ $SL(2,R)/U(1)$ coset theory. This is
rather different from the bosonic case where black hole theory  is
reached by a perturbation away from $c=1$ Liouville theory instead
of being equivalent to it.

\vskip 5mm

While this manuscript was in preparation, there appeared several
papers which studied black holes in type 0 theory
\cite{AdS2,type01,type02,type03}. Authors of Ref.~\cite{type02} in
particular mention possible role of Wilson line in formation of
nonextremal black holes in their system, which we suspect to be
related to the role of Wilson line in our bosonic case.

\vspace{1cm}

\noindent{\Large\bf Acknowledgements}

\vspace{5mm} \noindent We would like to thank Y.Michishita and
J.Raeymaekers for valuable discussions.

\end{document}